\documentclass[12pt]{article}
\usepackage{epsfig}
\newcommand{\noi}{\noindent}
\newcommand{\eq}{\begin{equation}}
\newcommand{\en}{\end{equation}}
\newcommand{\eqa}{\begin{eqnarray}}
\newcommand{\ena}{\end{eqnarray}}

\newcommand{\tr}{\mbox{Tr}\,}
\newcommand{\eff}{e \! f \! f}

\newcommand{\vx}{{\vec x}}

\newcommand{\bpartial}{{\bar \partial}}

\newcommand{\lra}{\longrightarrow}

\begin{document}
\hbox{}
\noindent June  1996 \hfill HU Berlin--EP--96/21
%

\vspace{0.5cm}
\begin{center}

\renewcommand{\thefootnote}{\fnsymbol{footnote}}
\setcounter{footnote}{0}

{\LARGE Gauge fixing, zero--momentum modes and the calculation
of masses on a lattice}
\footnote{Work supported by the Deutsche
Forschungsgemeinschaft under research grant Mu 932/1-4} \\

\vspace*{0.5cm}
{\large
V.K.~Mitrjushkin $\mbox{}^1$\footnote{Permanent adress:
Joint Institute for Nuclear Research, Dubna, Russia}
}\\

\vspace*{0.2cm}
{\normalsize
$\mbox{}^1$ {\em Institut f\"{u}r Physik, Humboldt-Universit\"{a}t,
10115 Berlin, Germany}\\     
}     
\vspace*{0.5cm}
{\bf Abstract}
\end{center}

It is shown that the zero--momentum modes can strongly affect the values
of the masses, for example the magnetic screening mass $m_m$, calculated
from gauge--dependent correlators with zero momentum.

\renewcommand{\thefootnote}{\arabic{footnote}}
\setcounter{footnote}{0}

\vspace{2cm}

The lattice approach gives the possibility to calculate numerically gauge
invariant objects without gauge fixing.  However in many practical
situations it is rather useful to calculate gauge dependent quantities.
For example, calculating the (gauge--variant) gluon correlators one 
can attempt to obtain information about gauge--invariant
observables, like energies and masses.

The usual way to proceed is the following \cite{mo}.
Define zero--momentum operators ${\cal O}_{\mu}(\tau )$:

\eq
{\cal O}_{\mu}(\tau ) = \sum_{\vx} {\cal O}_{\mu}(x);
\qquad {\cal O}_{\mu}(x) = \frac{1}{2i} \Bigl(U_{x\mu}-U_{x\mu}^{\dagger} \Bigr)~,
\en

\noi where $U_{x\mu} \in SU(N)$ are link fields. The symbol $~\tau$
means the separation in one of the four euclidian directions, and ${\vec x}$
corresponds to the three complementary directions.
Choose the Lorentz (or landau) gauge by maximizing the trace of the link fields
$~{\cal G} \equiv \sum_{x\mu} \tr 
\Bigl[ U_{x\mu} + U_{x\mu}^{\dagger} \Bigr]$,
and calculate the (connected) correlator

\eq
\Gamma_{\mu}(\tau ) = \left\langle \tr \Bigl( 
{\cal O}_{\mu}(\tau ){\cal O}_{\mu}(0) \Bigr) \right\rangle ~.
                    \label{correl_sun}
\en

\noi The effective masses ${\tilde m}(\tau )$ are defined as 
\eq
\frac{\cosh {\tilde m}(\tau ) (\tau +1-\frac{1}{2}N_4)}
{\cosh {\tilde m}(\tau ) (\tau -\frac{1}{2}N_4)}
= \frac{\Gamma_{\mu}(\tau +1)}{\Gamma_{\mu}(\tau )}~.
              \label{m_eff}
\en

\noi At zero temperature in the confinement phase the large--$\tau$
behaviour of ${\tilde m}(\tau )$ is
supposed to describe the gluon mass $m_g$ if $\tau$ is chosen along the
`temperature' direction $x_4$.  At high temperatures in the chromoplasma
phase the long distance behaviour of $\Gamma_{1;2}(z)$ and
$\Gamma_{4}(z)$ is expected to determine the magnetic and electric
screening masses $m_m$ and $m_e$, respectively (see, for example,
\cite{hkr}).

The translation--invariant gauge fixing, e.g.  Lorentz gauge
condition, does not exclude the appearence of the zero--momentum modes
in the periodic volume.  These modes are usually assumed to be
non-important.  Indeed, their contribution to the Wilson loops is rather
small, at least in the weak coupling region \cite{ka2}. 
It is the aim of this note to discuss the possible
influence of the zero--momentum mode on the gauge--dependent
(zero--momentum) correlators.

In what follows $N_{\mu}$ means the size of the lattice in the direction
$\mu$, and $V_4 =N_1 N_2 N_3 N_4$.
Boundary conditions are periodic, and the lattice spacing is chosen 
to be unity.
The lattice derivatives are:
$\partial_{\mu} f(x) = f(x+{\hat \mu}) -f(x)$ and  
$\bpartial_{\mu} f(x) = f(x) - f(x-{\hat \mu})$.

Throughout this paper the restriction to small quantum
fluctuations $A_{x\mu}$ about some constant value 
is implicitly presumed.

\vspace{0.25cm}

It is rather instructive to consider first the $U(1)$ gauge theory.  The
photon is massless in the Coulomb phase, and the perturbation theory is
expected to be reliable in the weak coupling limit.

In this case $~U_{x\mu} = \exp (i\theta_{x\mu})~$ with $~\theta_{x\mu}
\in (-\pi ;\pi ]$, and the Wilson action $~S(\theta )~$ and the
partition function $Z$ are

\eqa
S(\theta ) &=& \frac{1}{g^2} \sum_{x} \sum_{\mu \nu =1}^4
        \,  \bigl( 1 - \cos \theta_{x;\mu \nu} \bigr) ~;
\\
\nonumber \\
Z &=& \int \! [d\theta_{x \mu}] ~ e^{- S(\theta )}~,
\qquad [d\theta_{x \mu}]= \prod_{x\mu}\! \frac{d\theta_{x\mu}}{2\pi}~,
                                          \label{part_u1}
\ena

\noi where $~\theta_{x;\mu\nu}
= \partial_{\mu} \theta_{x\nu} -\partial_{\nu} \theta_{x\mu}$.
The action is invariant with respect to the gauge transformations
$~U_{x \mu} \stackrel{\Omega}{\lra}  U_{x \mu}^{\Omega} =
\Omega_x U_{x \mu} \Omega_{x+\mu}^{\dagger}~;
\quad \Omega_x \in U(1)~$.

 The standard perturbation approach reads as follows
(see, for example, \cite{mm}).
Define the Faddeev--Popov determinant $J(\theta )$ :

\eq
1 = J(\theta ) \int \! \prod_x \! d \Omega_x \, 
\prod_{x} \delta \Bigl( F_x(\theta^{\Omega} ) \Bigr)~,
                  \label{faddeev}
\en

\noi where $F_x(\theta )$ is some gauge fixing functional.
Setting it equal to

\eq
F_x(\theta ) = \sum_{\mu} \bpartial_{\mu} \sin \theta_{x\mu} - C_x~,
                 \label{func_F}
\en

\noi inserting the identity (\ref{faddeev}) in the integral of
eq.(\ref{part_u1}), multiplying both sides by
$\prod_x \exp \big( -\frac{1}{\alpha g^2} C_x^2 \big)~$
and integrating over $~C_x~$ one obtains

\eq
Z \sim \int \! [d\theta_{x\mu}]
 ~ e^{- S_{\eff}(\theta )}~; \quad S_{\eff}= S+S_{gf}+S_{FP} ~,
\en

\noi where $~S_{gf} =  \frac{1}{\alpha g^2} \sum_x 
\Bigl( \sum_{\mu} \bpartial_{\mu} \sin \theta_{x\mu} \Bigr)^2$,
and $S_{FP}$ is the contribution from the corresponding
Faddeev--Popov determinant\footnote{$S_{F\! P}~$ does not contribute
in the gaussian approximation.}.
The choice $~\alpha =0~$ corresponds to the Lorentz gauge.
The average of any (gauge invariant or not) functional $\Phi (\theta )$
is defined as

\eq
\langle \Phi \rangle = \frac{1}{Z} \int
\! [d\theta_{x\mu}] ~ \Phi (\theta ) \cdot e^{- S_{\eff}(\theta )}~.
                        \label{func_phi}
\en

\noi Making the substitution $~\theta_{x\mu} =g A_{x\mu}$ with
$~A_{x\mu} \in (-\frac{\pi}{g},\frac{\pi}{g}]$, expanding 
in powers of $~g^2~$ and extending the limits of integration to 
$~[-\infty ;\infty ]$
one recovers standard perturbation theory on a lattice
with a free action $S^{(0)}_{\eff}$ :

\eq
S^{(0)}_{\eff} = \sum_{xy}\sum_{\mu\nu} A_{x\mu}L^{\mu\nu}_{xy}A_{y\nu};
\qquad
L^{\mu\nu}_{xy} = \left[
\Bigl( 1-\frac{1}{\alpha} \Bigr) \bpartial_{\nu} \partial_{\mu} 
-\delta_{\mu\nu} \sum_{\rho} \bpartial_{\rho} \partial_{\rho} 
\right] \delta_{xy}~.
\en

\noi The matrix $L^{\mu\nu}_{xy}$ has a zero eigenvalue
with $x$--independent eigenstates  $\phi_{\mu}$.
Correspondingly, the free propagator is ill--defined at zero momentum.

The origin of the problem is quite general : if the solutions 
of the classical equations of motion depend on some continuum parameter
then the gaussian action has zero modes \cite{pol,ar}.
The perturbative scheme described above corresponds to 
the expansion about the constant solution  $\theta^{cl.}_{x\mu}=0$
of the classical equation of motion

\eq
\frac{\partial S (\theta )}{\partial \theta_{x \mu}} 
=\frac{2}{g^2} \sum_{\nu } \bpartial_{\nu}
\sin \theta_{x;\mu \nu} = 0~.
                  \label{classeq_u1}
\en

\noi However, there is a continuum of solutions
$\theta^{cl.}_{x\mu}=\phi_{\mu}$ with different $\phi_{\mu}$
\footnote{There are also non--constant solutions of the
eq.(\ref{classeq_u1}) which are not discussed in this paper (see
\cite{vkm}).}.  The chosen gauge condition -- Lorentz gauge -- does not
lift this degeneracy, and this produces the residual zero modes of the
quadratic form $S^{(0)}_{\eff}$.

The usual way to handle this problem is to exclude the $p=0$ mode from
all Feynman diagramms (see, e.g., \cite{hk}), which is justified in the
case of gauge--invariant objects (e.g. Wilson loops) \cite{ka2}.  
This is equivalent to the calculation of {\it another} average $\langle
\Phi \rangle^{\prime}$ :

\eq
\langle \Phi \rangle^{\prime} = \frac{1}{Z}
\int \! [d\theta_{x \mu}] ~ \delta \Bigl( \sum_x \theta_{x\mu} \Bigr)
\cdot \Phi (\theta ) \cdot e^{- S_{\eff}(\theta )}~.
                        \label{func_phi_prime}
\en

To calculate $\langle \Phi \rangle$ perturbatively one
should keep the zero--momentum mode under control
in the perturbation expansion.
It can be achieved by repeating the Faddeev--Popov trick 

\eq
1 = J_0\int \! [d\phi_{\mu}]~ 
\exp \left\{-\frac{1}{\epsilon}\sum_{\mu} 
\Bigl(\phi_{\mu} - \frac{1}{V_4} \sum_x\theta_{x\mu} \Bigr)^2 
\right\} \Big|_{\epsilon \to 0}~;
\qquad [\phi_{\mu}] = \prod_{\mu =1}^{4} \! d\phi_{\mu}~, 
\en

\noi and making the change of variables $~\theta_{x\mu} = \phi_{\mu}
+gA_{x\mu}$.  The average functional $\langle \Phi \rangle$ can be now
represented in the form

\eq
\langle \Phi \rangle \sim \int_{-\pi}^{\pi} \!
[d\phi_{\mu}] \! \int \! [dA_{x \mu}] 
\prod_{\mu=1}^4 \! \delta \Bigl( \sum_x A_{x\mu} \Bigr) 
\Phi (\phi +gA) e^{- S_{\eff}(\phi +gA)}
= \int_{-\pi}^{\pi} \! [d\phi_{\mu}] \langle \Phi \rangle (\phi ).
                        \label{func_phi_new}
\en

\noi In eq.(\ref{func_phi_new}) one can safely expand in powers of
$g^2$ and extend the limits of integration of $A_{x\mu}$ to
$[-\infty ;\infty ]$.
The zero--momentum modes are not gaussian, and the 
integration over $\phi_{\mu}$ in eq.(\ref{func_phi_new})
should stay compact.

The difference between the two kinds of averaging -- 
$~\langle \Phi \rangle^{\prime}~$ and $~\langle \Phi \rangle~$ --
manifests itself even in the lowest (gaussian) approximation.
Let us calculate the photon correlator

\eq
\Gamma_{\mu}(\tau ) = \left\langle 
{\cal O}_{\mu}(\tau ){\cal O}_{\mu}(0) \right\rangle;
\qquad
{\cal O}_{\mu}(\tau ) = \sum_{\vx} \sin \theta_{x\mu}.
                    \label{correl_u1}
\en

\noi Evidently, $\left\langle {\cal O}_{\mu}\right\rangle =0$.
 Ruling out the zero--momentum mode as in eq.(\ref{func_phi_prime})
one obtains for $\tau =x_4$

\eqa
\Gamma^{\prime}_i (\tau ) &=& \frac{V_3^2}{2V_4}
\sum_{p_4 \ne 0} \frac{e^{i\tau p_4}}{4\sin^2\frac{p_4}{2}}~;
\qquad i=1;2;3~;
                 \label{cor_nozm_u1}
\\
\Gamma^{\prime}_4 (\tau ) &=&0  \quad \mbox{at} \quad \alpha =0,
\nonumber
\ena

\noi where $V_3=N_1N_2N_3$.
The most important observation concerning $\Gamma^{\prime}_i (\tau )$
is that it is not zero. It has a rather non--trivial
dependence on $\tau$ and can even become negative.

Taking into account the zero--momentum modes $\phi_{\mu}$
one obtains in the Lorentz gauge

\eqa
\Gamma_i (\tau ;\phi ) &=& C^2 + \frac{V_3^2}{2V_4} \cos^2 \phi_i 
\sum_{p_4 \ne 0} \frac{e^{i\tau p_4}}{4\sin^2\frac{p_4}{2}}~;
                 \label{cor_zm_u1}
\\
\Gamma_4 (\tau ;\phi ) &=& C^2~;
\qquad C^2 = \frac{V_3^2}{g^2} \sin^2 \phi_i (1-bg^2)~,
\nonumber
\ena

\noi wheree $b \simeq 0.058$.
The non--zero value of $\Gamma_4 (\tau )$ is the indicator
of the presence of zero--momentum modes. 

\begin{figure}[htb]
\vspace{-1.0cm}
\begin{center}
%
%
\epsfig{file=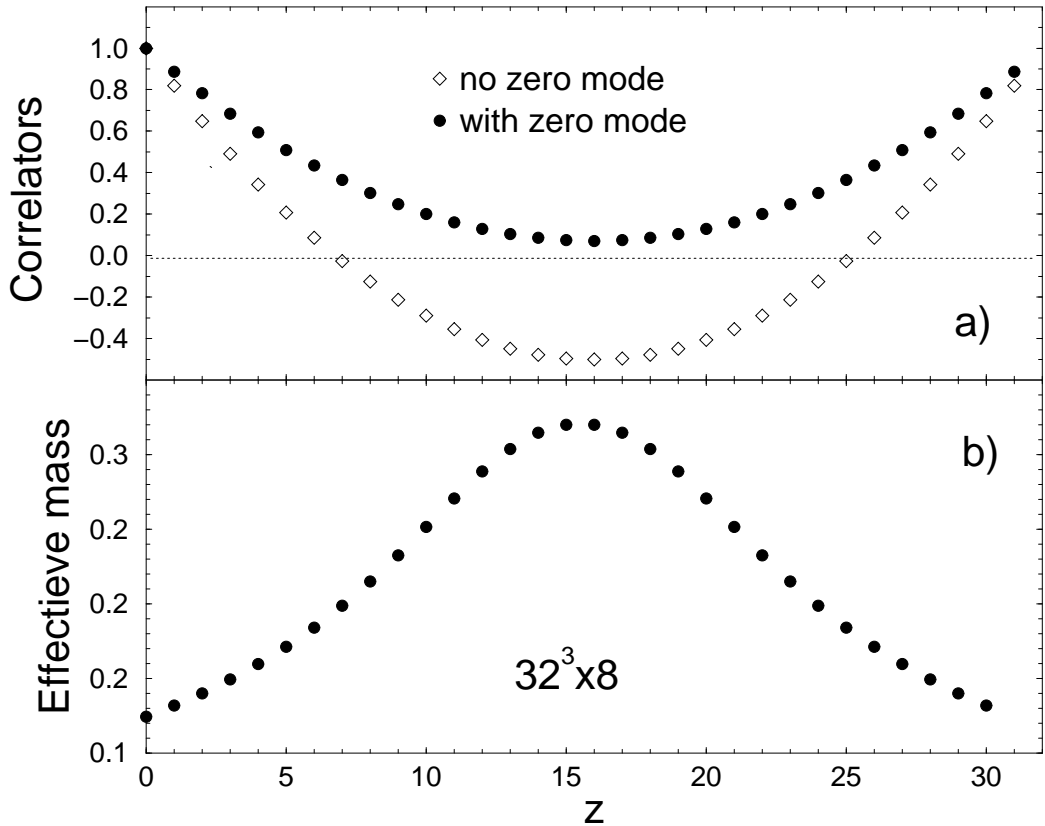,width=14cm}
\end{center}
\vskip 0.25truecm
\caption{
The correlators $\Gamma_1^{\, \prime}(z)$ and $\Gamma_1(z)$
({\bf a}) and the effective ('magnetic') mass ${\tilde m}(z)$ ({\bf b}) 
on a $32^3\times 8$ lattice. The zero--momentum mode $\phi_1$ was 
chosen to be $\phi_1 =0.01$, and $g^2=1$.
}
\label{fig:cor_1}
\vskip -0.2truecm
\end{figure}

The $\tau$--dependence of the correlator $\Gamma_i(\tau )$ 
can mimic the non--zero photon mass ${\tilde m}(\tau )$
{\it if} defined in accordance with eq.(\ref{m_eff}).
As an example, in Figure \ref{fig:cor_1}a one can see two correlators
$\Gamma_1^{\prime}(z)$ and $\Gamma_1(z)$ with separation along
the space--like direction
$z=x_3$ calculated on a $32^3\times 8$ lattice.  The
zero--momentum mode was chosen to be $\phi_1 =0.01$ and $g^2=1$.
The effective mass ${\tilde m}(z)$ corresponding to the 
correlator $\Gamma_1(z;\phi )$ is shown in Figure \ref{fig:cor_1}b. 
Of course, one can hardly expect the appearence of the nonzero 
magnetic mass
in the Coulomb phase in the pure gauge $U(1)$ theory.
It is rather a `phantom' produced by a) the non--trivial $\tau$--dependence
of the correlator $\Gamma^{\prime}(\tau )$ and b) the zero--momentum modes.

%
%
%
%

\vspace{0.25cm}

The extension to the case of the non--abelian theories is
straightforward.  As well as in the abelian case, there is a degeneracy
of the $x$--independent solutions $U^{cl.}_{\mu}$ of the classical
equations of motion due to the toroidal structure of the periodic
lattice \cite{ka1,ka2}.  These solutions have zero action if $~\Bigl[
U^{cl.}_{\mu},U^{cl.}_{\nu} \Bigr] =0$, they are called torons.  An
example of a toron is $U^{cl.}_{\mu} = \exp\{ i\phi_{\mu} T\}$, where
$T$ is one of the generators of the gauge group and $\phi_{\mu}$ are
four numbers.  The perturbative expansion deals with fluctuations about
the torons.

Suppose for simplicity that $U^{cl.}_{\mu} =\exp\{ i\phi_{\mu}T \}$ is a
diagonal matrix.  One can represent the link field in the form
$~U_{x\mu} =\exp\{ iA_{x\mu} \} \exp\{ i\phi_{\mu}T\}~$ where
$~A_{x\mu}~$ satisfy the condition $~\sum_x A_{x\mu}^{diag.}=0$.  The
non--abelian field $~A_{x\mu} = {\vec A}_{x\mu}{\vec T}~$ has the
non--diagonal zero--momentum modes $~{\bar A}_{\mu} = A_{\mu}(p=0)$, and
their contribution to the effective action in the lowest approximation
is

\eq
S^{(0;\, zero~modes)}_{\eff} \sim \sum_{\mu\nu} \tr
\left\{\Bigl( D_{\mu}{\bar A}_{\nu}-D_{\nu} {\bar A}_{\mu}\Bigr)^2 \right\},
\en

\noi where $D_{\mu}A_{\nu}=
\exp\{ i\phi_{\mu}T \} A_{\nu} \exp\{ -i\phi_{\mu}T \} -A_{\nu}$.
In principle, torons are not physically equivalent. 
If $~U^{cl.}_{\mu} \in Z_N~$ (singular toron) then $~D_{\mu}{\bar A}_{\nu} =0$,
and the zero--momentum modes are not gaussian, as in the $U(1)$ case.
Careful consideration shows that the zero--momentum corrections to
the Wilson loops are rather small \cite{ka2}.

On the contrary, the dependence of the gauge--variant correlators
$\Gamma_{\mu}(\tau )$ on the zero--momentum can be very strong.  The
chosen gauge--fixing does not prohibit the fluctuations about some
constant value $\phi^a_{\mu}$, and the shift $~A^a_{x\mu} \to A^a_{x\mu}
+ \phi^a_{\mu}~$ (for small fields) produces the appearence of the
non--negative constant term in the correlator as it happens in the
$U(1)$ theory (compare eq.(\ref{cor_nozm_u1}) and eq.(\ref{cor_zm_u1})).

\vspace{0.5cm}

The main conclusion is that the contribution of the zero--momentum modes
to the zero--momentum gauge--variant correlators cannot be ignored.  Even
rather small zero--momentum modes ($|\phi_{\mu}| \ll 1$) can become very
competitive when compared with the `normal' modes, affecting strongly the
effective masses.

It would be interesting to know how the zero--momentum modes
`develop' during the updating procedure, and what their
dependence on the chosen gauge is.
These questions deserve a furthur study.

{}From the practical point of view, presumably, it is preferable
to generate configurations in MC calculations suppressing the
zero--momentum modes as in eq.(\ref{func_phi_prime}).

\vspace{1cm}
\noi {\large {\bf Acknowledgements}}

\vspace{0.25cm}
\noi I would like to express my gratitude to U. Heller for useful
discussions.

\vspace{1cm}

\end{document}